\magnification=\magstep1 
\font\bigbfont=cmbx10 scaled\magstep1
\font\bigifont=cmti10 scaled\magstep1
\font\bigrfont=cmr10 scaled\magstep1
\vsize = 23.5 truecm
\hsize = 15.5 truecm
\hoffset = .2truein
\baselineskip = 14 truept
\overfullrule = 0pt
\parskip = 3 truept
\def\frac#1#2{{#1\over#2}}

\def\fS{{\sl{S}}}

\def\bbbc{{\mathchoice {\setbox0=\hbox{$\displaystyle\rm C$}\hbox{\hbox
to0pt{\kern0.4\wd0\vrule height0.9\ht0\hss}\box0}}
{\setbox0=\hbox{$\textstyle\rm C$}\hbox{\hbox
to0pt{\kern0.4\wd0\vrule height0.9\ht0\hss}\box0}}
{\setbox0=\hbox{$\scriptstyle\rm C$}\hbox{\hbox
to0pt{\kern0.4\wd0\vrule height0.9\ht0\hss}\box0}}
{\setbox0=\hbox{$\scriptscriptstyle\rm C$}\hbox{\hbox
to0pt{\kern0.4\wd0\vrule height0.9\ht0\hss}\box0}}}}
\def\ket#1{|#1\rangle}
\def\bra#1{\langle #1|}

\nopagenumbers
\topinsert
\vskip 3.2 truecm
\endinsert
\centerline{\bigbfont WHAT IS A COOPER PAIR?$^*$} 
\vskip 20 truept
\centerline{\bigifont G. Ortiz$^{1}$ and J. Dukelsky$^2$}
\vskip 8 truept
\centerline{\bigrfont $^1$Theoretical Division, Los Alamos National Laboratory} 
\vskip 2 truept
\centerline{\bigrfont Los Alamos, New Mexico 87545, USA}
\vskip 8 truept
\centerline{\bigrfont $^2$Instituto de Estructura de la Materia, CSIC}
\vskip 2 truept
\centerline{\bigrfont Serrano 123, 28006 Madrid, Spain} 
\vskip 1.8 truecm

\centerline{\bf 1.  INTRODUCTION}
\vskip 12 truept


Superconductivity and Fermi superfluidity have been subjects of 
increasing interest since their discovery. It seems as if nature wisely
entangled Bose-Einstein (BE) condensation with these collective
phenomena which are direct macroscopic manifestations of quantum
physics. Recently, the nature of Cooper pairs in the BCS-BEC crossover
has regained attention due to the observation of a large fraction of
preformed fermion pairs on the BCS side of the Feshbach resonance in
ultracold atomic Fermi gases [1]. While several theoretical
explanations were proposed [2], the interpretations are still
controversial. The root of the controversy is understanding what
represents a Cooper pair in a correlated Fermi system, a concept that
has been extensively used in the literature to describe the properties
and dynamics of the superconducting state but without a unique and
precise definition.

Although the question {\it What is a Cooper pair?} seems academic, it
is indeed a question of  upmost practical importance given the great
advances we are witnessing in ultracold atom physics and quantum
information. New experimental probes allow access to unusual regimes of
strongly-coupled systems with unprecedented degree of control. 
Therefore, concepts need to be sharply defined to correctly interpret
these new experimental results. The truth is that much of our current
physical intuition is built upon mean-field solutions. Can one get new
useful insight out of exactly-solvable models? This manuscript
discusses the Cooper-pair concept at the very fundamental level and
proposes a qualitatively different scenario of the BCS-BEC crossover,
based on the exact solution to the BCS Hamiltonian [3]. Only in the
extreme BEC limit does this new scenario and Leggett's {\it naive}
ansatz [4] become identical. As we will see, the Cooper pair in a
correlated Fermi system is merely a statistical concept whose
manifestation and interpretation depend upon the particular measurement
or physical operation performed in the system.

While the superconducting and Fermi superfluid thermodynamic states
represent a mixed-system of quasifree and pair-correlated fermions, the
molecular BEC which arises in the dilute and strong coupling limit has
all fermions  {\it bound} into pairs forming a unique macroscopic
quantum state. It is by now well accepted in which sense these states
represent a Bose-Einstein (BE) condensation. What defines a BE
condensation in an interacting $N$-particle system is spontaneous gauge
symmetry breaking, or phase coherence, of its quantum state (a concept
that strictly applies in the thermodynamic limit (TL)). Yang [5]
provided a detailed mathematical characterization of this phenomenon
which manifests itself as off-diagonal long-range order (ODLRO) or,
equivalently, by the existence of an eigenvalue of order $N$ in a
reduced density matrix $\hat{\rho}_n$, defined as
$\hat{\rho}_n(kl\cdots|ij\cdots)={\rm Tr}[a_ka_l \cdots \hat{\rho}
\cdots a^\dagger_ja^\dagger_i]$ ($i,j,k,l,\cdots$, labelling
single-particle states annihilated by bosonic and/or fermionic
operators $a$'s) where $n$ is the number of particles forming the {\it
smallest unit} or {\it complex} that condenses. ($n=2$ electrons for
BCS superconductors).

\vskip 28 truept
\centerline{\bf 2.  WHAT IS THE MANY-BODY WAVEFUNCTION?}
\vskip 12 truept

The above definition of a BEC does not imply the naive view of a
many-body state of quantum objects with identical internal
wavefunctions. Indeed, we will argue that a current-carrying  ground
state (GS) of a uniform superconducting or Fermi superfluid
$N$-particle system is of the form
$$
\Psi(x_1,\cdots,x_N)=  {\cal A} \left [ \phi_1(x_1, x_2) \cdots
\phi_{N/2}(x_{N-1}, x_N) \right ] \ , 
\eqno(1)
$$
with $x_j=({\bf r}_j, \sigma_j)$, antisymmetrizer ${\cal A}$, and the {\it
pair} state
$$
\phi_{\alpha}(x_i,x_j)=\
e^{i\bf{q}\cdot\frac{(\bf{r}_i+\bf{r}_j)}{2}} \
\varphi_\alpha(\bf{r}_i-\bf{r}_j)  \
\chi(\sigma_i,\sigma_j) \ , 
$$
where $\chi$ is a spin function of a certain symmetry, $\bf{q}$ is the
pair center-of-mass momentum, and $\varphi_\alpha(\bf{r})$ the internal
wavefunction which may represent either a quasimolecular resonant state
or a scattering state  (i.e., a mixed state), depending upon the
strength of the interaction between particles. The antisymmetrizer 
${\cal A}$ is applied to the particle index $x_i$. For example,
$\Psi(x_1,x_2,\cdots,x_N)=-\Psi(x_2,x_1,\cdots,x_N)$. 

For pedagogical reasons we will recall the one-Cooper-pair problem and
then address the question of what happens when we add more pairs
($\bf{q}=\bf{0}$ in the GS). The Cooper-pair solution can be
recovered from the wavefunction (1) by assuming that $N-2$ fermions
$c^\dagger_{{\bf k} \sigma}$ are free, filling the lowest momentum states
${\bf k}$ (of energy $\varepsilon_{\bf k}$) up to the Fermi momentum
${\bf k}_F$,
while {\it only} an additional fermion pair (with momenta $k> k_F$)
feels an attractive $s$-wave interaction in the spin singlet channel
$$
\ket \Psi_{\rm C}=\sum_{k>k_{F}}\frac{1}{2\varepsilon_{\bf k}-E}
c_{{\bf k}\uparrow}^{\dagger}c_{-{\bf k}\downarrow}^{\dagger} \ \ket F \ .
\eqno(2)
$$
The role of the Fermi sea, $\ket F=\prod_{k\leq k_F} c_{{\bf
k}\uparrow}^{\dagger}c_{-{\bf k}\downarrow}^{\dagger}\ket 0$, is to
Pauli-block states below the Fermi energy $\varepsilon_F$. Assuming
that the attractive pairing interaction $G<0$ is constant around the
Fermi energy, the eigenvalue $E$ satisfies
$$
\frac{1}{G}=\sum_{k>k_F} \frac{1}{E-2 \varepsilon_{\bf k}} \ ,
$$ 
which  always has a real negative solution, implying that the Cooper
pair is bound regardless of the strength of the attractive interaction.
The Fermi sea is then unstable against the formation of {\it bound
pairs} of electrons.

What happens when the pairing interaction {\it also} affects
electrons in the Fermi sea? The answer to this question is the BCS
theory whose canonical form considers a simple variational state of
the form (1) with all {\it identical} internal
wavefunctions (a pair mean-field)
$$
\hat{P}_M \ket {{\rm BCS}} =\left(  \Lambda^{\dagger} \right)^{M} \ket 0
\ , \ \Lambda^{\dagger}=\sum_{\bf k}\frac{v_{\bf k}}{u_{\bf k}}
c_{{\bf k}\uparrow}^{\dagger} c_{-{\bf k}\downarrow}^{\dagger} \ ,
$$
where $\hat{P}_M$ is the projector onto the state with $M$ fermion
pairs, and $v_{\bf k}$, $u_{\bf k}$ are the BCS amplitudes
$$
\pmatrix{v_{\bf k}^2\cr u_{\bf k}^2 \cr}=\frac{1}{2}\left
(1\mp\frac{\varepsilon_{\bf k}-\mu}{
\sqrt{(\varepsilon_{\bf k}-\mu)^2+\Delta^2}} \right )
$$
with $\Delta$ the gap and $\mu$ the chemical potential. In the language
of first quantization this mean-field state represents a Pfaffian state:
Given the skew-symmetric matrix ($\phi_{ij}=\phi(x_i, x_j))$
$$
\hat{\Psi}=\pmatrix{
0 & \phi_{12} & \phi_{13} & \cdots & \phi_{1N} \cr
-\phi_{12} & 0& \phi_{23} & \cdots & \phi_{2N} \cr
-\phi_{13} & -\phi_{23}& 0 & \cdots & \phi_{3N} \cr
\vdots& \vdots&  \vdots   & \cdots & \vdots \cr
-\phi_{1N}&-\phi_{2N}&-\phi_{3N} & \cdots& 0
} \ ,
$$
${\sl Pf} \hat{\Psi}=\Psi_{\rm BCS}(x_1,\cdots,x_N)=  {\cal A} \left [ 
\phi(x_1, x_2) \cdots \phi(x_{N-1}, x_N) \right ]$, which represents a
linear combination of $(N-1)!!$ terms.

Since the structure of the BCS pair is averaged over the whole system,
it does not characterize a Cooper pair in the
superconducting/superfluid medium except in the extreme strong-coupling
and dilute limits where all pairs are {\it bounded} and
non-overlapping. (This approximate state is not the exact ground state
of the reduced BCS Hamiltonian, Eq. (3).) Sometimes, the structure of
the Cooper pair is associated with the pair-correlation function $\bra
{{\rm BCS}} c_{{\bf k}\uparrow}^{\dagger}c_{-{\bf k}\downarrow}^{\dagger
}\ket {{\rm BCS}} =u_{\bf k}v_{\bf k}$ describing the pair correlation
among fermions of the same pair as well as the exchange between
fermions of different pairs.

What is the nature of a Cooper pair in a Fermi superfluid or
superconducting medium? To address this question we will  use the exact
solution to the reduced BCS Hamiltonian
$$
H =\sum_{\bf{k}}\varepsilon_{\bf{k}}\
{n}_{\bf{k}}+\frac{G}{V}\sum_{\bf{ k,k^{\prime }}}\
c_{\bf{k\uparrow }}^{\dagger }c_{-\bf{k\downarrow }
}^{\dagger }c_{-\bf{k^{\prime }\downarrow
}}^{\;}c_{\bf{k^{\prime }\uparrow }}^{\;}  \ ,
\eqno(3)
$$
proposed by Richardson 40 years ago [3,6]. $H$ involves all terms with
time-reversed pairs (${\bf k}\uparrow,-{\bf k}\downarrow$) from a
contact interaction. It is consistent with an effective single-channel
description of the BCS-BEC crossover theory [4] in terms of a
zero-range potential. Realistic finite-range interactions produce
qualitatively similar results along the crossover [7]. The main
differences are expected in the BEC region where the reduced BCS
Hamiltonian, Eq. (3), cannot account for the reduction in the
scattering length of the composite pairs [8] and the corresponding
depletion of the molecular condensate [9].

For simplicity we will consider $N=N_\uparrow+N_\downarrow$ spin-1/2
(i.e., 2-flavor) fermions in a three-dimensional box of volume $V$ with
periodic boundary conditions, interacting through an attractive
constant ($s$-wave-singlet-pairing) potential. (Other pairing
symmetries can also be accommodated.) Exact $N=2M+\nu$ particle
eigenstates of $H$ can be written as
$$
\ket {\Psi}=\prod_{\ell=1}^M \fS^+_\ell \ket {\nu} \ , {\rm with }
\ \fS^+_\ell=\sum_{\bf k} \frac{1}{2 \varepsilon_{\bf k} -E_\ell}
c_{\bf{k}\uparrow}^{\dagger}
c_{-\bf{k}\downarrow}^{\dagger} ,
\eqno(4)
$$
where $\left\vert \nu\right\rangle \equiv$ $\left\vert
\nu_{1},\nu_{2}\cdots,\nu_{L}\right\rangle$ is a state of $\nu$
unpaired fermions ($\nu=\sum_{\bf k}\nu_{\bf k}$, with $\nu_{\bf
k}=1,0$) defined by $c_{-\bf{k\downarrow}}^{\;}c_{\bf{k\uparrow}}^{\;}
\left\vert \nu\right\rangle =0$, and $n_{\bf k} \left\vert
\nu\right\rangle =\nu_{\bf k}\left\vert \nu\right\rangle$. $L$ is the
total number of single particle states. The GS $\ket{\Psi_0}$ is in the
$\nu=0$ ($N_\uparrow\!=\!N_\downarrow$) sector.

Each eigenstate $\ket{\Psi}$ is completely defined by a set of $M$ (in
general, complex) spectral parameters (pair energies) $E_\ell$ which
are a solution of the Richardson's equations
$$
1+\frac{G}{V}\sum_{\bf k}\frac{1-\nu_{\bf k}}{2\varepsilon_{\bf k}-E_\ell}+
\frac{2G}{V}\sum_{m \left( \neq \ell \right) =1}^{M}\frac {1}{
E_{\ell}-E_{m}}=0 \ , \eqno(5)
$$
and the eigenvalues of $H$ are ${\cal E}=\sum_{\bf k} \varepsilon_{\bf
k} \nu_{\bf k}+\sum_{\ell=1}^{M}E_{\ell}$ [6,10]. A crucial observation
is that if a complex $E_\ell$ satisfies (5), its complex-conjugate
$E^*_\ell$ is also a solution. Thus, $\ket{\Psi}$ restores
time-reversal invariance. The ansatz (4) is a natural generalization of
the Cooper-pair problem without an inert Fermi sea, and with all pairs
subjected to the pairing interaction. The pair structure in (4) is 
similar to the Cooper pair in (2), and the many-body state has the form
(1) with all pairs {\it different}. In a sense, that will become clear
later on, the many-body state (4) represents a statistical distribution
of {\it resonances} (Schafroth's intuition of the superconducting state). 

\vskip 28 truept
\centerline{\bf 3.  THERMODYNAMIC LIMIT OF RICHARDSON'S EQUATIONS}
\vskip 12 truept

Since we are concerned with uniform bulk Fermi systems, we are
interested in the TL (i.e., $N, V \rightarrow \infty$ with
$\rho=N/V=k_F^3/3\pi^2$ = constant) of the Richardson's exact solution
Eq. (5). This limit was studied by Gaudin [11] when the energy spectrum
$\Omega \in [-\omega,\omega]$ is bounded, and $\nu=0$. Eqs. (5) reduce
to the well-known {\it gap equation}
$$
\frac{1}{2}\int_{\Omega}d\varepsilon ~\frac{g(\varepsilon)}{\sqrt{
\left( \varepsilon -\mu\right) ^{2}+\Delta^{2}}}+\frac{1}{G}=0 \ ,
\eqno(6)
$$
where $g(\varepsilon)$ represents the density of states. In our case
$\Omega \in  [0,\infty]$ is unbounded with
$g(\varepsilon)=m^{3/2}\varepsilon^{1/2}/(\sqrt{2} \pi^{2}\hbar^{3})$
for $\varepsilon_{\bf k}=\hbar^2k^2/2m$. Due to the absence of an upper
cutoff, Eq. (6) is singular and the TL becomes a subtle mathematical
procedure whose solution will be presented here.

There are two ways to regularize the problem: One can either introduce
an energy cutoff in the integrals or one can cancel the singularities
with physical quantities whose bare counterpart diverges in the same
way [4]. For this problem, Eq. (3), the physical quantity is the
scattering length $a_s$ given by
$$
\frac{m}{4\pi \hbar^2
a_s}=\frac{1}{G}+\frac{1}{2}\int_{0}^{\infty} d\varepsilon
~\frac{g(\varepsilon)}{\varepsilon} \ .
$$
The non-singular gap equation (after integration [13]) is
$$
\frac{1}{k_F a_s}=\eta = (\mu^{2}+\Delta ^{2})^{1/4} \
P_{\frac{1}{2}}\left( -\frac{\mu}{\sqrt{\mu^{2} +\Delta ^{2}}}\right)
\ , \eqno(7)
$$
where energies are now in units of $\varepsilon_F=\hbar^2k_F^2/2m$ and
lengths in units of $\xi_F=1/k_F$. $P_\beta(x)$ is the Legendre
function of the first kind of degree $\beta$. The equations for the
conservation of the number of pairs $M$
$$
-\frac{4}{3\pi} =\eta \mu  +  (\mu^{2}+\Delta ^{2})^{3/4} \
P_{\frac{3}{2}}\left(  -\frac{\mu}{\sqrt{\mu^{2}+\Delta
^{2}}}\right) , \eqno(8)
$$
and GS energy density [12], for arbitrary strength $\eta$,
$$
{\cal E}_0=\frac{1}{V} \frac{\bra{\Psi_0} H
\ket{\Psi_0}}{\langle \Psi_0 | \Psi_0 \rangle}
=-\frac{k_F^3}{20\pi} \left [ \frac{\eta \Delta^2}{2} -
\frac{4}{\pi} \mu \right ] \ ,
\eqno(9)
$$
do not need regularization, these are {\it exact} results. Indeed, for
a given $\eta$, the chemical potential $\mu$ and gap $\Delta$ need to
be determined self-consistently from Eqs. (7) and (8) (Fig. 1). Then,
the GS energy can be computed as a function of density using Eq. (9).
It shows no phase segregation. The {\it exact} binding (or
condensation) energy per electron pair ${\cal E}_B$  is 
$$
{\cal E}_B=-\frac{2}{\rho} \left [ {\cal E}_0 -
\frac{1}{V}\frac{\bra{F} H \ket{F}} {\langle F | F \rangle} \right
]=-\frac{2}{\rho} \left [ {\cal E}_0- \frac{k_F^3}{5\pi^2} \right ]
=\frac{3 \pi}{10} \left [\frac{\eta \Delta^2}{2}- \frac{4}{\pi}
(\mu-1) \right ] \ .
$$

One can analytically determine the asymptotic $\eta \rightarrow \pm
\infty$ behavior of $\mu$ and $\Delta$ by properly combining Eqs. (7)
and (8), and making use of the defining Eqs. for the Legendre functions
$P_\beta(x)$. Indeed, one can convert  Eqs. (7) and (8) into coupled
differential equations, with one particular combination leading to
$$
\frac{\partial \ln(\mu^2+\Delta^2)}{\partial \eta}=\left ( \frac{\pi
\Delta}{2}\right)^2 \frac{\eta}{1+ \left ( \frac{\pi
\Delta}{2}\right)^2\frac{\eta^2}{4}} \ .
$$
The asymptotics can be determined by using this equation and the results
are depicted in Table 1.

\topinsert
\centerline{\bf{Table 1}}
\vskip 12 truept
\noindent
Analytic expressions for selected values of $\eta=1/k_Fa_s$;
$x$ is the root of $P_{\frac{1}{2}}$, i.e. $P_{\frac{1}{2}}(x)=0$, and
$E(y)$ is the complete elliptic integral of the second kind.
Note that $5\pi^2{\cal E}_0(\eta=0)/k_F^3=\mu(\eta=0)\approx 0.590606$.
\vskip 27 truept

%
\newbox\hdbox%
\newcount\hdrows%
\newcount\multispancount%
\newcount\ncase%
\newcount\ncols
\newcount\nrows%
\newcount\nspan%
\newcount\ntemp%
\newdimen\hdsize%
\newdimen\newhdsize%
\newdimen\parasize%
\newdimen\spreadwidth%
\newdimen\thicksize%
\newdimen\thinsize%
\newdimen\tablewidth%
\newif\ifcentertables%
\newif\ifendsize%
\newif\iffirstrow%
\newif\iftableinfo%
\newtoks\dbt%
\newtoks\hdtks%
\newtoks\savetks%
\newtoks\tableLETtokens%
\newtoks\tabletokens%
\newtoks\widthspec%
%
%
\immediate\write15{%
CP SMSG GJMSINK TEXTABLE --> TABLE MACROS V. 851121 JOB = \jobname%
}%
%
%
\tableinfotrue%
\catcode`\@=11
%
%
\def\tstrut{\vrule height3.1ex depth1.2ex width0pt}%
\def\and{\char`\&}
\def\tablerule{\noalign{\hrule height\thinsize depth0pt}}%
\thicksize=1.5pt
\thinsize=0.6pt
\def\thickrule{\noalign{\hrule height\thicksize depth0pt}}%
\def\ctr#1{\hfil\ #1\hfil}%
%
%
%
%
\tablewidth=-\maxdimen%
\spreadwidth=-\maxdimen%
\def\tabskipglue{0pt plus 1fil minus 1fil}%
%
%
\centertablestrue%
%
%
%
%
\parasize=4in%
\gdef\ARGS{########}
\gdef\headerARGS{####}
\def\@mpersand{&}
{\catcode`\|=13
\gdef\letbarzero{\let|0}
\gdef\letbartab{\def|{&&}}%
\gdef\letvbbar{\let\vb|}%
}
{\catcode`\&=4
\def\ampskip{&\omit\hfil&}
\catcode`\&=13
\let&0
\xdef\letampskip{\def&{\ampskip}}%
\gdef\letnovbamp{\let\novb&\let\tab&}
}
\def\begintable{
   \begingroup%
   \catcode`\|=13\letbartab\letvbbar%
   \catcode`\&=13\letampskip\letnovbamp%
   \def\multispan##1{
      \omit \mscount##1%
      \multiply\mscount\tw@\advance\mscount\m@ne%
      \loop\ifnum\mscount>\@ne \sp@n\repeat%
   }
   \def\|{%
      &\omit\widevline&%
   }%
   \ruledtable
}
\long\def\ruledtable#1\endtable{%
%
%
%
   \offinterlineskip
   \tabskip 0pt
   \def\widevline{\vrule width\thicksize}
   \def\endrow{\@mpersand\omit\hfil\crnorm\@mpersand}%
   \def\crthick{\@mpersand\crnorm\thickrule\@mpersand}%
   \def\crnorule{\@mpersand\crnorm\@mpersand}%
   \let\nr=\crnorule
   \def\endtable{\@mpersand\crnorm\thickrule}%
   \let\crnorm=\cr
%
%
   \edef\cr{\@mpersand\crnorm\tablerule\@mpersand}%
   \the\tableLETtokens
%
%
   \tabletokens={&#1}
%
%
   \countROWS\tabletokens\into\nrows%
   \countCOLS\tabletokens\into\ncols%
%
%
   \advance\ncols by -1%
   \divide\ncols by 2%
   \advance\nrows by 1%
%
%
   \iftableinfo %
      \immediate\write16{[Nrows=\the\nrows, Ncols=\the\ncols]}%
   \fi%
%
%
   \ifcentertables
      \ifhmode \par\fi
      \line{
      \hss
   \else %
      \hbox{%
   \fi
      \vbox{%
         \makePREAMBLE{\the\ncols}
         \edef\next{\preamble}
         \let\preamble=\next
         \makeTABLE{\preamble}{\tabletokens}
      }
      \ifcentertables \hss}\else }\fi
   \endgroup
   \tablewidth=-\maxdimen
   \spreadwidth=-\maxdimen
}
\def\makeTABLE#1#2{
   {
   \let\ifmath0
   \let\header0
   \let\multispan0
%
%
   \ncase=0%
   \ifdim\tablewidth>-\maxdimen \ncase=1\fi%
   \ifdim\spreadwidth>-\maxdimen \ncase=2\fi%
   \relax
%
   \ifcase\ncase %
      \widthspec={}%
   \or %
      \widthspec=\expandafter{\expandafter t\expandafter o%
                 \the\tablewidth}%
   \else %
      \widthspec=\expandafter{\expandafter s\expandafter p\expandafter r%
                 \expandafter e\expandafter a\expandafter d%
                 \the\spreadwidth}%
   \fi %
   \xdef\next{
      \halign\the\widthspec{%
      #1
      \noalign{\hrule height\thicksize depth0pt}
      \the#2\endtable
%
      }
   }
   }
   \next
}
\def\makePREAMBLE#1{
   \ncols=#1
   \begingroup
   \let\ARGS=0
   \edef\xtp{\widevline\ARGS\tabskip\tabskipglue%
   &\ctr{\ARGS}\tstrut}
   \advance\ncols by -1
   \loop
      \ifnum\ncols>0 %
      \advance\ncols by -1%
      \edef\xtp{\xtp&\vrule width\thinsize\ARGS&\ctr{\ARGS}}%
   \repeat
   \xdef\preamble{\xtp&\widevline\ARGS\tabskip0pt%
   \crnorm}
   \endgroup
}
\def\countROWS#1\into#2{
   \let\countREGISTER=#2%
   \countREGISTER=0%
   \expandafter\ROWcount\the#1\endcount%
}%
\def\ROWcount{%
   \afterassignment\subROWcount\let\next= %
}%
\def\subROWcount{%
   \ifx\next\endcount %
      \let\next=\relax%
   \else%
      \ncase=0%
      \ifx\next\cr %
         \global\advance\countREGISTER by 1%
         \ncase=0%
      \fi%
      \ifx\next\endrow %
         \global\advance\countREGISTER by 1%
         \ncase=0%
      \fi%
      \ifx\next\crthick %
         \global\advance\countREGISTER by 1%
         \ncase=0%
      \fi%
      \ifx\next\crnorule %
         \global\advance\countREGISTER by 1%
         \ncase=0%
      \fi%
      \ifx\next\header %
         \ncase=1%
      \fi%
      \relax%
      \ifcase\ncase %
         \let\next\ROWcount%
      \or %
         \let\next\argROWskip%
      \else %
      \fi%
   \fi%
   \next%
}
\def\counthdROWS#1\into#2{%
\dvr{10}%
   \let\countREGISTER=#2%
   \countREGISTER=0%
\dvr{11}%
\dvr{13}%
   \expandafter\hdROWcount\the#1\endcount%
\dvr{12}%
}%
\def\hdROWcount{%
   \afterassignment\subhdROWcount\let\next= %
}%
\def\subhdROWcount{%
   \ifx\next\endcount %
      \let\next=\relax%
   \else%
      \ncase=0%
      \ifx\next\cr %
         \global\advance\countREGISTER by 1%
         \ncase=0%
      \fi%
      \ifx\next\endrow %
         \global\advance\countREGISTER by 1%
         \ncase=0%
      \fi%
      \ifx\next\crthick %
         \global\advance\countREGISTER by 1%
         \ncase=0%
      \fi%
      \ifx\next\crnorule %
         \global\advance\countREGISTER by 1%
         \ncase=0%
      \fi%
      \ifx\next\header %
         \ncase=1%
      \fi%
\relax%
      \ifcase\ncase %
         \let\next\hdROWcount%
      \or%
         \let\next\arghdROWskip%
      \else %
      \fi%
   \fi%
   \next%
}%
{\catcode`\|=13\letbartab
\gdef\countCOLS#1\into#2{%
   \let\countREGISTER=#2%
   \global\countREGISTER=0%
   \global\multispancount=0%
   \global\firstrowtrue
   \expandafter\COLcount\the#1\endcount%
   \global\advance\countREGISTER by 3%
   \global\advance\countREGISTER by -\multispancount
}%
\gdef\COLcount{%
   \afterassignment\subCOLcount\let\next= %
}%
{\catcode`\&=13%
\gdef\subCOLcount{%
   \ifx\next\endcount %
      \let\next=\relax%
   \else%
      \ncase=0%
      \iffirstrow
         \ifx\next& %
            \global\advance\countREGISTER by 2%
            \ncase=0%
         \fi%
         \ifx\next\span %
            \global\advance\countREGISTER by 1%
            \ncase=0%
         \fi%
         \ifx\next| %
            \global\advance\countREGISTER by 2%
            \ncase=0%
         \fi
         \ifx\next\|
            \global\advance\countREGISTER by 2%
            \ncase=0%
         \fi
         \ifx\next\multispan
            \ncase=1%
            \global\advance\multispancount by 1%
         \fi
         \ifx\next\header
            \ncase=2%
         \fi
         \ifx\next\cr       \global\firstrowfalse \fi
         \ifx\next\endrow   \global\firstrowfalse \fi
         \ifx\next\crthick  \global\firstrowfalse \fi
         \ifx\next\crnorule \global\firstrowfalse \fi
      \fi
\relax
      \ifcase\ncase %
         \let\next\COLcount%
      \or %
         \let\next\spancount%
      \or %
         \let\next\argCOLskip%
      \else %
      \fi %
   \fi%
   \next%
}%
\gdef\argROWskip#1{%
   \let\next\ROWcount \next%
}
\gdef\arghdROWskip#1{%
   \let\next\ROWcount \next%
}
\gdef\argCOLskip#1{%
   \let\next\COLcount \next%
}
}
}
\def\spancount#1{
   \nspan=#1\multiply\nspan by 2\advance\nspan by -1%
   \global\advance \countREGISTER by \nspan
   \let\next\COLcount \next}%
\def\dvr#1{\relax}%
\def\header#1{%
\dvr{1}{\let\cr=\@mpersand%
\hdtks={#1}%
\counthdROWS\hdtks\into\hdrows%
\advance\hdrows by 1%
\ifnum\hdrows=0 \hdrows=1 \fi%
\dvr{5}\makehdPREAMBLE{\the\hdrows}%
\dvr{6}\getHDdimen{#1}%
{\parindent=0pt\hsize=\hdsize{\let\ifmath0%
\xdef\next{\valign{\headerpreamble #1\crnorm}}}\dvr{7}\next\dvr{8}%
}%
}\dvr{2}}
\def\makehdPREAMBLE#1{
\dvr{3}%
\hdrows=#1
{
\let\headerARGS=0%
\let\cr=\crnorm%
\edef\xtp{\vfil\hfil\hbox{\headerARGS}\hfil\vfil}%
\advance\hdrows by -1
\loop
\ifnum\hdrows>0%
\advance\hdrows by -1%
\edef\xtp{\xtp&\vfil\hfil\hbox{\headerARGS}\hfil\vfil}%
\repeat%
\xdef\headerpreamble{\xtp\crcr}%
}
\dvr{4}}
\def\getHDdimen#1{%
\hdsize=0pt%
\getsize#1\cr\end\cr%
}
\def\getsize#1\cr{%
\endsizefalse\savetks={#1}%
\expandafter\lookend\the\savetks\cr%
\relax \ifendsize \let\next\relax \else%
\setbox\hdbox=\hbox{#1}\newhdsize=1.0\wd\hdbox%
\ifdim\newhdsize>\hdsize \hdsize=\newhdsize \fi%
\let\next\getsize \fi%
\next%
}%
\def\lookend{\afterassignment\sublookend\let\looknext= }%
\def\sublookend{\relax%
\ifx\looknext\cr %
\let\looknext\relax \else %
   \relax
   \ifx\looknext\end \global\endsizetrue \fi%
   \let\looknext=\lookend%
    \fi \looknext%
}%
%
%
\def\tablelet#1{%
   \tableLETtokens=\expandafter{\the\tableLETtokens #1}%
}%
\catcode`\@=12

\nrows= 5
\ncols= 5
\begintable
$\eta$ ||& $\mu$ |& $\Delta$ |& ${\cal E}_B$ |&
$\lambda$ 
\cr
$\rightarrow - \infty$ || & $1+(\frac{\pi}{8}\eta
-\frac{5}{8}) \Delta^2$ |& $\frac{8}{e^{2}} \exp{\frac{\pi
\eta}{2}}$ |& $\frac{3}{4}\Delta^2$ |& $\frac{3\pi}{e^{2}} \exp{\frac{\pi
\eta}{2}}$
\cr
$0$ ||& -$\frac{x}{E\left( \frac{1-x}{2}\right)^{\frac{2}{3}}}$ |&
$\frac{\sqrt{1-x^2}}{E\left( \frac{1-x}{2}\right)^{\frac{2}{3}}}$ |&
$\frac{6}{5} \left ( 1+ \frac{x}{E\left(
\frac{1-x}{2}\right)^{\frac{2}{3}}}\right )$ |& $\frac{3\pi}{8\sqrt{2}}
\frac{\sqrt{1+x}(1-x)}{E\left(\frac{1-x}{2}\right)}$
\cr
$\frac{8\pi^{2/3}}{\Gamma[\frac{1}{4}]^{8/3}}$ ||& 0 |&
$\sqrt{2\eta}$ |& $\frac{6}{5} \left ( 1+ \frac{\pi
\eta^2}{4}\right )$ |& $\frac{3\pi^{3/2}}{\sqrt{2}\Gamma[\frac{1}{4}]^{2}}$
\cr
$\rightarrow + \infty$ ||& $-\eta^2$ |&
$\sqrt{\frac{16\eta}{3\pi}}$ |& $2 \eta^2$| & 1
\endtable
\vskip 14 truept
\endinsert

The complete solution of Eqs. (5) in the TL [11] amounts to determining
(for a given $\eta$) the set of pair energies $E_\ell$ which, for the
GS,  form a single open, continuous, and reflection-symmetric arc
$\Gamma$ with extreme points $E_F=2(\mu \pm i \Delta)$. Here, we simply
present the results without derivation. The equation for $\Gamma$ is
$$
\eqalignno{
0={\sl Re} \left[ \int_{0}^{\infty }d\varepsilon
~\sqrt{\varepsilon }  \left(  \frac{z+\left( \varepsilon -\mu
\right)\ln \left[ \frac{\left( E -\mu \right) +z}{i \Delta
}\right]}{\sqrt{\left( \varepsilon -\mu \right) ^{2}+\Delta^{2}}} -
\quad \quad \quad \quad \quad \quad \quad \quad \quad \quad \quad \quad 
\right .\right . \cr
 \left . \left .  \ln \left[ \frac{ \Delta
^{2}+\left( E -\mu \right) \left( \varepsilon -\mu \right) +z \
\sqrt{\left( \varepsilon -\mu \right) ^{2}+\Delta ^{2}}}{ i\Delta
\left( \varepsilon -E \right)} \right] \right ) \right] \ ,
&  \quad \ \  (10)
}
$$
where $z=\sqrt{\left( E -\mu \right) ^{2}+\Delta ^{2}}$.

\topinsert
\input psfig.sty
\centerline{\hskip0mm
\psfig{figure=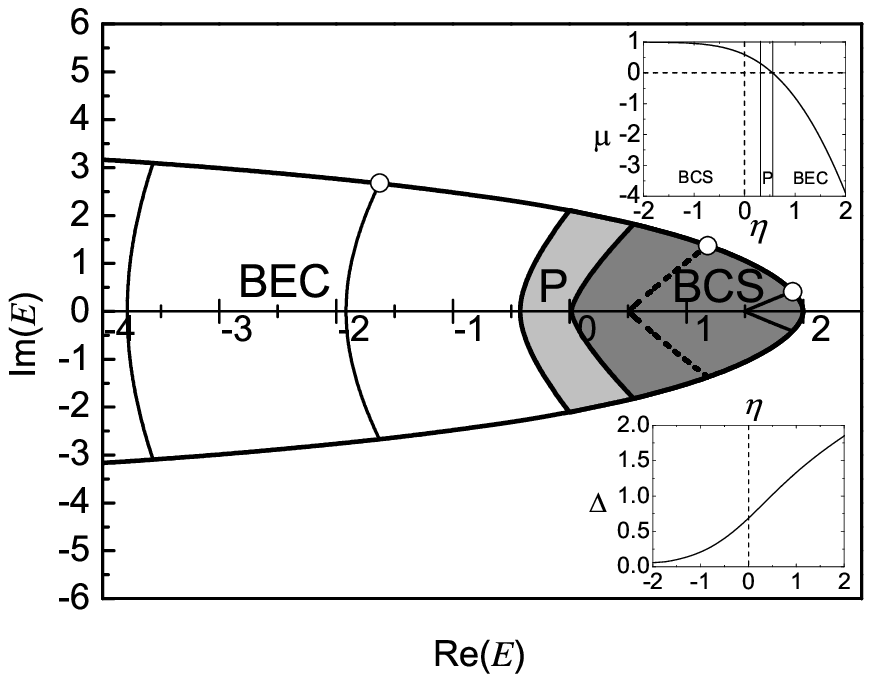,height=10truecm,width=11truecm,angle=0}}
\vskip -0.0truecm 
\noindent
{\bf Figure 1.} 
Ground state crossover diagram displaying the different regimes as a
function of $\eta$. A few arcs $\Gamma$, from Eq. (10), corresponding
to the values of $\eta=$ -1, 0, 0.37, 0.55, 1 and 2, whose extremes are
$E_F=2(\mu\pm i\Delta)$, are displayed. The dashed line corresponds to
the Feshbach resonance $\eta=0$. Open circles represent the values of
$E$ for which Cooper-pair wavefunctions are plotted in Fig. 3. Insets
show the behavior of the gap $\Delta$ and chemical potential $\mu$.
\vskip 6truept
\endinsert

\vskip 28 truept
\centerline{\bf 4.  BCS-TO-BEC CROSSOVER PHYSICS}
\vskip 12 truept

Figure 1 shows the BCS-BEC crossover diagram and several arcs
corresponding to particular values of $\eta$. Since a crossover diagram
is not a phase diagram, i.e. there is no symmetry order parameter or
non-analyticities sharply differentiating the regions, boundaries are
in principle arbitrary.

Here we have adopted the following criteria: The geometry of the arcs
$\Gamma$ serves us to establish a criterium for defining boundary
regions in the crossover diagram. In the extreme weak-coupling limit,
$\eta \rightarrow -\infty$, the pair energies are twice the single
particle energies ($E_\ell \rightarrow 2 \varepsilon_{\bf k}$). As soon
as the interaction is switched on, i.e. $k_Fa_s$ is an infinitesimal
negative number, a fraction $ {\rm f}$  of the pair energies close to
$2\mu$ become complex, forming an arc $\Gamma$ in the complex plane.
The fraction $1-{\rm f}$ of fermion pairs below the crossing energy $2
\varepsilon_c$ of $\Gamma$ with the real axis have real pair energies.
They are still in a sea of uncorrelated pairs with an effective Fermi
energy $\varepsilon_c$.  The dark grey region labelled BCS, which
extends from  $\eta=-\infty$ to $\eta=0.37$, is characterized by a
mixture of  complex pair energies with a positive real part and real
pair energies. $\eta=0.37$ is the value at which all pair energies are
complex, i.e. ${\rm f}=1$, and the effective Fermi sea has disappeared
($\varepsilon_c=0$). Within the BCS region we plotted the arcs for
$\eta=-1$ (solid line), with a fraction  ${\rm f}=0.35$ and an
effective Fermi energy $\varepsilon_c=0.747$ , and $\eta=0$ (dashed
line) with ${\rm f}=0.87$ and $\varepsilon_c=0.25$. The {\it pseudogap}
region P, indicated in light grey, extends from $\eta=0.37$ to
$\eta=0.55$ where $\mu=0$. Within this region the real part of the pair
energies changes from positive to negative, and P describes a mixture
of Cooper resonances and quasibound molecules. The BEC (white) region,
$\eta>0.55$, is characterized by all pair energies having negative real
parts, i.e. all pairs are quasibound molecules. Within the BEC region
we plotted two arcs with $\eta=1$ and $\eta=2$. As $\eta$ increases
further, $\Gamma$ tends to an almost vertical line with ${\sl
Re}(E)\sim 2\mu$, and $ -2\Delta \leq  {\sl Im}(E) \leq 2\Delta$.

\vskip 28 truept
\centerline{\bf 5.  WHAT IS THE STRUCTURE OF A COOPER PAIR?}
\vskip 12 truept

Having established a qualitative BCS-BEC crossover diagram, we turn now
to the nature of the Cooper pairs. If the literature is not clear about
the size $\xi$ of a Cooper pair (all lenghts are written in units of
$\xi_F=1/k_F$), it is partly because it is not clear what a Cooper pair
is. Pippard, in his nonlocal electrodynamics of the superconducting
state, introduced the characteristic length $\xi_0$ by using an
uncertainty-principle argument involving only electrons within a shell
of width 2$\Delta$ around the Fermi energy. The resulting quantity,
named Pippard's coherence length, $\xi_0=2/(\pi \Delta)$, is usually
associated to $\xi$. On the other hand, an acceptable definition could
be
$\xi=\sqrt{\bra{\varphi} r^2 \ket{\varphi}} \ , \ \langle
\varphi | \varphi \rangle =1$.
From Eq. (4), the Cooper-pair wavefunction is 
$$
\varphi_E({\bf r})=\frac{1}{V}\sum_{\bf{k}} \varphi_{\bf k}^E \
e^{i\bf{k}\cdot\bf{r}}=
A \ \frac{e^{-r\sqrt{-E/2}}}{r} \ , 
$$
with $A^2={\sl Im}(\sqrt{E/2})/2\pi \xi_F^3$, $\varphi_{\bf
k}^E=C/(2\varepsilon_{\bf k}-E)$, and $C$ being a normalization
constant. Thus,
$$
\xi_E=\frac{1}{{\sl Im}(\sqrt{E})} 
$$
for arbitrary interaction strength $\eta$. In the weak-coupling BCS
limit ($\Delta \ll \mu \approx 1$), when $E=E_F$, we get
$\xi_E=\pi\xi_0/\sqrt{2}$. On the other hand, in the same limit, if one
uses $\varphi_{\bf k}^{\rm P}= C_{\rm P} u_{\bf k} v_{\bf k}$, one gets
$\xi_{\rm P}=\xi_E/2$, and if one uses $\varphi_{\bf k}^{\rm BCS}=
C_{\rm BCS} v_{\bf k}/u_{\bf k}$, one gets $\xi_{\rm BCS}=\sqrt{21/2}$.

\topinsert
\input psfig.sty
\centerline{\hskip0mm
\psfig{figure=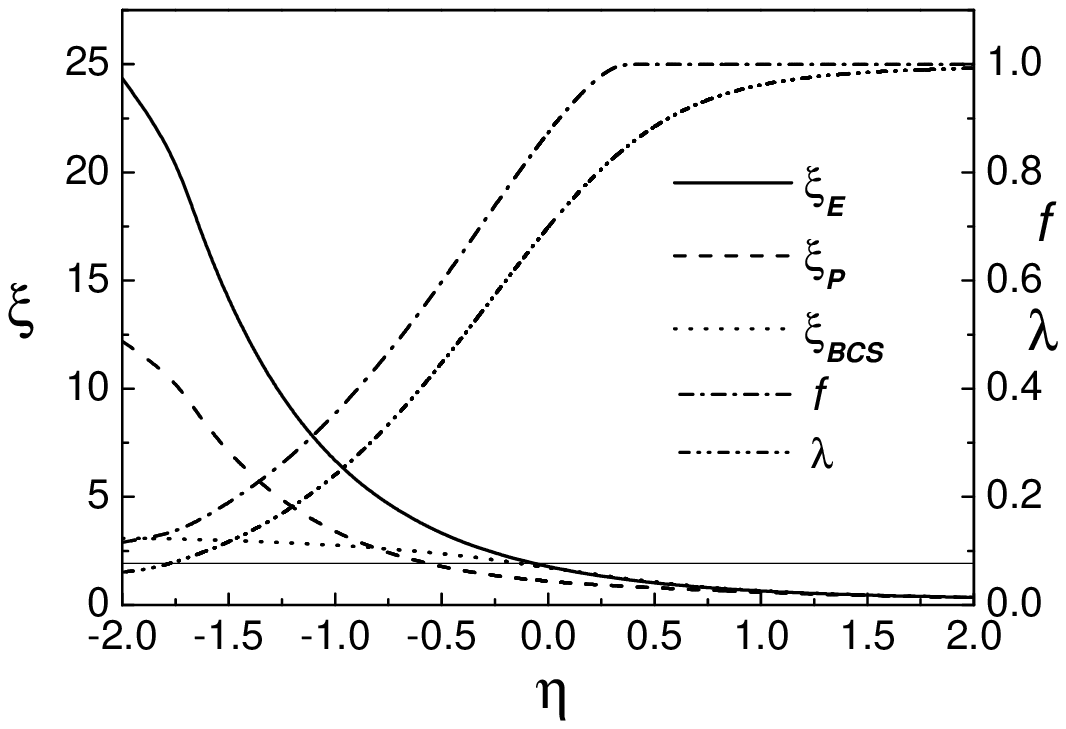,height=10truecm,width=11truecm,angle=0}}
\vskip -0.0truecm 
\noindent
{\bf Figure 2.} 
Different definitions of Cooper-pair sizes $\xi$, and condensate
fraction $\lambda$ and ${\rm f}$ as functions of the interaction
strength $\eta$. The horizontal solid line represents the interparticle
distance $r_s=(9\pi/4)^{1/3}$ (lenghts are written in units of
$\xi_F=1/k_F$). $\xi_E$ and $\xi_{\rm P}$ increase for  negative $\eta$
values, and eventually diverge when $\eta \rightarrow -\infty$ as the
inverse of the gap, like Pippard's coherence length $\xi_0=2/(\pi
\Delta)$.
\vskip 6truept
\endinsert

Analytic expressions (using dimensional regularization) for
$\xi_{\rm P}$ and $\xi_{\rm BCS}$ can also be obtained for
arbitrary coupling strength. The results are
$$
\xi_{\rm P}= \frac{1}{{\sl Im} \left( \sqrt{E_F}\right )} \sqrt{
1 - \frac{3}{4} \left ( 1 - \frac{({\sl Im} \left( \sqrt{E_F}\right
))^2}{|E_F|}\right )} \ , \ {\rm and }
$$
$$
\xi_{\rm BCS}= \frac{1}{\sqrt{|E_F|}}\sqrt{ \frac{42 \left (\gamma \ P_{
\frac{3}{2}} \left (\gamma \right )- P_{ \frac{1}{2}}\left ( \gamma
\right ) \right )-\frac{35}{|E_F|^{5/2}} \ {\sl Im}(E_F)  \ {\sl
Re}(E_F^{3/2})}{ \left (\gamma^2-5 \right ) \ P_{ \frac{3}{2}} \left (
\gamma \right )+  4\gamma \ P_{ \frac{1}{2}}\left ( \gamma \right )}} \
,
$$
where $\gamma=-{\sl Re}(E_F)/|E_F|$. Figure 2 displays the different
sizes, all labelled by $\xi$, as a function of $\eta$. Clearly, in (4)
there is no unique pair size $\xi_E$ but a distribution of sizes.
Although in  the BCS region there is a distribution of sizes from the
smallest pair with $E=E_F$ in the extremes of the arc to an almost
infinite size pairs when they are close to the real axis, already at
resonance ($\eta=0$) most of the pairs have the same structure. This
latter feature becomes more pronounced in the BEC region. Figure 2
shows in solid line the smallest size in $\Gamma$ corresponding to
$E_F$. On the other hand, there is a unique pair size for $\varphi_{\rm
P}$ and $\varphi_{\rm BCS}$; the dashed and dotted lines correspond to
the $\xi_{\rm P}$ and $\xi_{\rm BCS}$ sizes respectively. Notice the
relation between sizes and the interparticle distance
$r_s=(9\pi/4)^{1/3}$ which is indicated as a full horizontal line.
While $\xi_E$ and $\xi_{\rm P}$ increase for negative $\eta$ values,
and eventually diverge when $\eta \rightarrow -\infty$,  $\xi_{\rm
BCS}$ tends to a constant value, related to the interparticle distance,
showing its essentially uncorrelated nature. The fact that $\xi_{\rm P}
< \xi_E/2$ in the overlapping BCS region is a clear demonstration that
$\xi_{\rm P}$ measures the mean distance between time-reversed pairs
irrespective of the Cooper pair they belong to. For non-overlapping
pairs (BEC region) both sizes converge to the same values.

The various definitions of Cooper-pair wavefunctions are depicted in
Fig. 3, which compares $\varphi_{\rm BCS}(r)$, $\varphi_{\rm P}(r)$,
and $\varphi_{E}(r)$ for interaction strengths which correspond to the
BCS, Feshbach resonance, and BEC regions of Fig. 1. Notice that while
$\varphi_{\rm BCS}$, and $\varphi_{\rm P}$ are highly oscillatory in
the weak-coupling region, this is not the case with $\varphi_{E}$ which
always has an exponential form. Clearly, a single and unified picture
emerges for the crossover when using a many-body state such as (1):
$\varphi_{E}$ evolves smoothly through the crossover as it should. It
is important to mention that the three wavefunctions are {\it exactly}
the same in the strong coupling limit $\eta \rightarrow +\infty$. In
this limit the {\it naive} ansatz of Leggett [4] and the GS (1)
coincide, becoming the simple Pfaffian state $\Psi_{\rm
BCS}(x_1,\cdots,x_N)=  {\cal A} \left [  \phi(x_1, x_2) \cdots
\phi(x_{N-1}, x_N) \right ]$.


\vskip 28 truept
\centerline{\bf 6.  CONDENSATE FRACTION OF A FERMI SUPERFLUID}
\vskip 12 truept

The condensate fraction is a thermodynamic property charaterizing the 
superfuid/superconducting state. In an ideal Bose gas it is the
fraction of bosons occupying the lowest-energy quantum state. In a
nonideal (Bose or Fermi) gas, one cannot use this definition.
Following Yang [5], ODLRO in
$\hat{\rho}_2({\bf r}_1\!\!\uparrow\!{\bf r}_2\!\!\downarrow \!\!| \, {\bf
r}_3\!\!\uparrow\!{\bf r}_4\!\!\downarrow) \rightarrow \phi^*({\bf
r}_1\!\!\uparrow,{\bf r}_2\!\!\downarrow)  \phi({\bf
r}_3\!\!\uparrow,{\bf r}_4\!\!\downarrow)$
may be used to define
$$ 
\lambda =
\frac{2}{N} \!\!\int \!\!d {\bf r}_1 d {\bf r}_2\ | \phi({\bf
r}_1\!\!\uparrow,{\bf r}_2\!\!\downarrow)|^2=\frac{3 \pi}{16}
\frac{\Delta^2}{{\sl Im}(\sqrt{\mu+ i \Delta})}
$$
as a measurement of the condensate fraction in a two-flavor Fermi
system. Figure 2 shows $\lambda$ and the fraction ${\rm f}$ of (Cooper)
pairs in the arc, that is, the fraction of complex spectral parameters.
Although $\lambda$ can qualitatively describe the fraction of
correlated pairs, the fraction ${\rm f}$ gives the value $1$ at the
BCS-pseudogap boundary ($\eta=0.37$) while $\lambda=1$ for $\eta
\rightarrow \infty$. We note that at resonance ($\eta=0$), ${\rm
f}=87\%$ of the fermions form Cooper pairs ($\lambda \approx 0.7$).
These results provide a qualitative explanation of the experiments in
[1]  without resorting to a projection method [2]. Close to resonance
on the BCS side, a fraction ${\rm f}\sim 80\%$ of the atoms form Cooper
pairs with sizes comparable to $r_s$. Those atom pairs are efficiently
transformed into {\it quasimolecules} during a rapid magnetic field
ramping across the resonance giving rise to the molecular condensate
fractions observed experimentally.


\topinsert
\input psfig.sty
\centerline{\hskip0mm
\psfig{figure=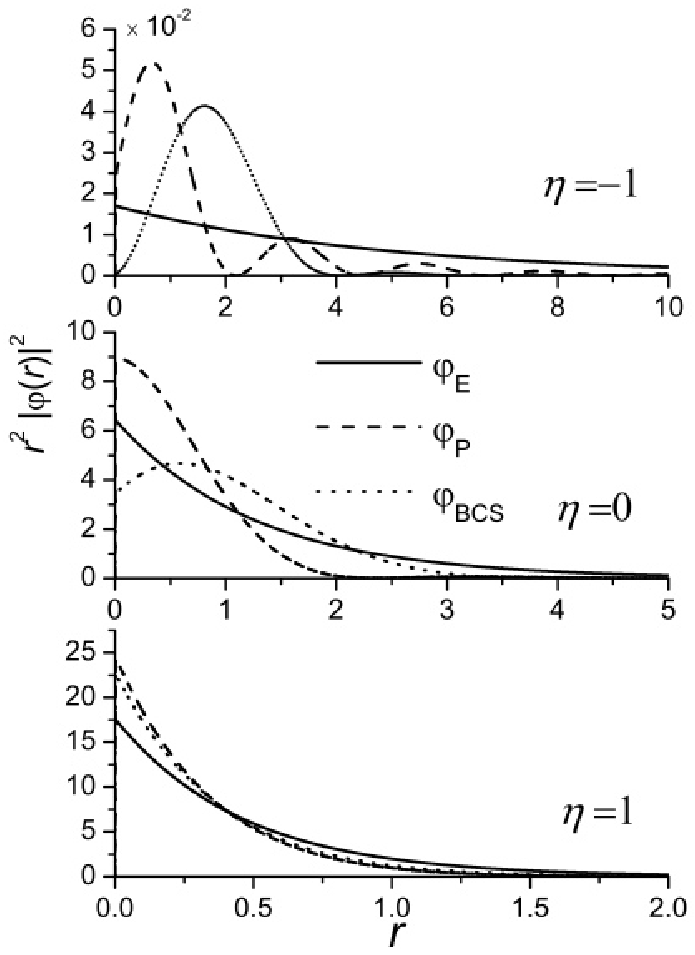,height=12truecm,width=9truecm,angle=0}}
\vskip -0.0truecm 
\noindent
{\bf Figure 3.} 
Cooper-pair wavefunctions for different $\eta$. The upper and middle
(Feshbach resonance) panels correspond to the BCS region, while the
bottom one is in the BEC region. Except for the Cooper $\varphi_E$
case, the other two wavefunctions always vanish at $r=0$. It is only in
the limit $\eta \rightarrow + \infty$ that the three states {\it
exactly} coincide.
\vskip 6truept
\endinsert

Finally, Table 1 summarizes various analytical results for
the chemical potential, gap, binding (or condensation) energy, and
condensate fraction, in four important limits of the single-channel
model: $\eta \rightarrow -\infty$ (BCS), $\eta=0$ (Resonance), $\mu=0$
(BCS-to-BEC crossover), and $\eta \rightarrow \infty$ (Extreme BEC
limit).

\vskip 28 truept
\centerline{\bf 7.  CONCLUSIONS}
\vskip 12 truept

In summary, we studied the BCS-BEC crossover problem, as well as the
nature of Cooper pairs in a correlated Fermi system, from the exact GS
$\ket{\Psi_0}$ of the reduced BCS Hamiltonian. We have analytically
determined its exact TL for the quadratic single-particle dispersion,
and calculated several quantities of physical interest, such as the
binding (or condensation) energy and the condensate fraction. The
fundamental difference between the BCS state and $\ket{\Psi_0}$ is the
fact that the former, being a mean-field, pairs off all electrons in a
unique paired wavefunction while the exact state contains a statistical
distribution of quasimolecular resonant and scattering states  (i.e., a
mixed state) that depends upon the strength of the interaction between
particles. It is only in the extreme BEC regime that the two pictures
coincide asymptotically. Then, what a Cooper pair is depends upon the
particular measurement or operation performed in the physical system. 
The validity of the present description is not restricted to integrable
pairing Hamiltonians, but rather the ansatz $\ket {\Psi_0}$, which  is a
natural extension of the Cooper problem, could be used as a variational
state for more realistic interactions. The Cooper-pair wavefunction
$\varphi_E$ has a fixed functional $s$-wave form that accommodates
pair-correlated fermions close to the Fermi energy in the weak coupling
limit as well as quasibound molecules in the BEC limit. It is free from
the characteristic oscillations displayed by $\varphi_{\rm BCS}$ and
$\varphi_{\rm P}$ related to exchange among pairs. The BCS region in the
crossover diagram represents a mixture of Cooper pairs and quasifree
fermions. These properties resemble the Schafroth description of the 
superconducting state [16].
It may be argued that the single-channel model is insufficient to
describe the system close to resonance where the degrees of freedom
associated to the molecular closed channel should be explicitly
incorporated. A derivation analogous to the one presented here can be
pursued by using a recently proposed atom-molecule integrable model
[14] which captures the essential features of the two-channel model
[15]. The structure of the Cooper pair in this new model is the same as
in (4).

\vskip 28 truept

\centerline{\bf REFERENCES}
\vskip 12 truept

$\!\!\!\!^*$ A longer version of G. Ortiz and J. Dukelsky, Phys. Rev. A 72, 043611 (2005). 

\item{[1]}
C. A. Regal, M. Greiner, and D. S. Jin, Phys. Rev. Lett.
{\bf 92}, 040403 (2004); M. W. Zwierlein {\it et. al}, Phys. Rev.
Lett. {\bf 92}, 120403 (2004); Phys. Rev. Lett. {\bf 94}, 180401
(2005).

\item{[2]}
R. B. Diener, and T-L. Ho,  cond-mat/0404517; A.
Perali, P. Pieri, and G. C. Strinati, cond-mat/0501631.

\item{[3]}
R. W. Richardson, Phys. Lett. {\bf 3}, 277 (1963);
Nucl. Phys. {\bf 52}, 221 (1964).

\item{[4]}
D. M. Eagles, Phys. Rev. {\bf 186}, 456 (1969); A. J.
Leggett, in {\it Modern trends in the theory of condensed matter},
edited by A. Pekalski and R. Przystawa (Springer Verlag, Berlin, 1980);
P. Nozi\`eres and S. Schmitt-Rink, J. Low Temp. Phys. {\bf 59}, 195 (1985);
J. R. Engelbrecht, M. Randeria, and C. A. R. S\'a de Melo, Phys. Rev. B
{\bf 55}, 15153 (1997).

\item{[5]}
C. N. Yang, Rev. Mod. Phys. {\bf 34}, 694 (1962).


\item{[6]}
G. Ortiz {\it et al.}, Nucl. Phys. B {\bf 707}, 421
(2005).

\item{[7]}
M. M. Parish {\it et al.}, Phys.Rev. B {\bf 71}, 064513
(2005).

\item{[8]}
 D. S. Petrov, C. Salomon and G. V. Shlyapnikov,
Phys. Rev. Lett. {\bf 93}, 090404 (2004).

\item{[9]}
G. E. Astrakharchik {\it et al.},
cond-mat/0507483.

\item{[10]}
J. Dukelsky {\it et al.}, Rev. Mod.
Phys. {\bf 76}, 643 (2004).

\item{[11]}
M. Gaudin, {\it Mod\`eles exactement r\'esolus}
(Les Editions de Physique, Les Ulis, 1995), p. 261; J. M. Roman,
G. Sierra, and J. Dukelsky, Nucl. Phys. B {\bf 634}, 483 (2002).

\item{[12]}
N. N. Bogoliubov, Physica {\bf 26}, S1 (1960), proved
that the GS energy per particle of (3) in the TL equals
the mean-field BCS result. (We verified that the complete set of
integrals of motion [6] are also equal in the TL.) There is
no existing proof that GSs $\hat{P}_M \ket {\rm BCS}$ and (4) are
the same; this is a subtle issue. As variational states they are
clearly different. An open question is what
observables distinguish them at the minimum of (3) in
the TL.

\item{[13]}
M. Marini {\it et al.}, Eur.
Phys. J. B {\bf 1}, 151 (1998); T. Papenbrock and G. F. Bertsch,
Phys. Rev. C {\bf 59}, 2052 (1999).


\item{[14]}
J. Dukelsky, {\it et. al}, Phys. Rev. Lett. {\bf 93}, 050403
(2004).

\item{[15]}
M. Holland, {\it et. al}, Phys. Rev. Lett. {\bf 87},
120406 (2001); E. Timmermans {\it et. al}, Phys. Lett. A {\bf 285},
228 (2001).

\item{[16]}
J. M. Blatt, {\it Theory of superconductivity}
(Academic Press, New York, 1964).

\vskip 28 truept

\centerline{\bf APPENDIX: USEFUL INTEGRALS}
\vskip 12 truept

We simply quote the results of some integrals used in this article. Some
of them were obtained using the technique of dimensional regularization 
($z=a+ib$, $z^\ast=a-ib$, $\gamma=-a/|z|$, excluding ${\sl Re}(z)=a\ge
0 \ {\rm and} \ {\sl Im}(z)=b=0$) 

$$
\eqalignno{
&\int_{0}^{\infty}\!\!\! dk\frac{k^{2}}{\left( k^{2}-z\right) \left(
k^{2}-z^{\ast }\right) }=i\frac{\pi}{2\left(
\sqrt{z}-\sqrt{z^{\ast}}\right) } \ . 
\cr
&\int_0^\infty\!\!\!  d k \ \frac{k^4}{(k^{2}-z)^2(k^{2}-z^{\ast})^2}=
\frac{\pi}{32} \frac{1}{({\sl Im}
\left( \sqrt{z}\right ))^3} \ . 
\cr
&\int_0^\infty \!\!\! d k \ \frac{k^4 (k^2 -
a)^2}{(k^{2}-z)^3(k^{2}-z^{\ast})^3}= \frac{\pi}{32} \frac{1}{({\sl Im}
\left( \sqrt{z}\right ))^3} \left [ 1 - \frac{3}{4} \left ( 1 -
\frac{({\sl Im} \left( \sqrt{z}\right ))^2}{|z|}\right ) \right ]  \ .
\cr
&\int_{0}^{\infty}\!\!\! dk \ \sqrt{k} \  (\sqrt{(k-z)(k-z^{\ast})}-(k-a) )^2
=\frac{4 \pi}{35} \sqrt{|z|^7} \left ( \left (\gamma^2-5 \right ) \ P_{
\frac{3}{2}} \left ( \gamma \right )+  4\gamma \ P_{ \frac{1}{2}}\left
( \gamma \right ) \right ) \ .
\cr
&\int_0^\infty \!\!\! d k \ \frac{k^4 \left ( (k^2 -a)-
\sqrt{(k^{2}-z)(k^{2}-z^{\ast})} \right )^2}{\pi
\sqrt{|z|^5} \ (k^{2}-z)(k^{2}-z^{\ast})} = \frac{3}{5} 
(\gamma \ P_{ \frac{3}{2}} \left (\gamma \right )- P_{
\frac{1}{2}}\left ( \gamma \right ) )-\frac{b}{2}  \
\frac{{\sl Re}(z^{3/2})}{\sqrt{|z|^5}}\ .
\cr
&\int_0^\infty \!\!\! dk \ \sqrt{k} \left [
\frac{1}{\sqrt{(k-z)(k-z^{\ast})}}-\frac{1}{k}\right]= - \pi
\sqrt{|z|} \ P_{\frac{1}{2}} \left (
\gamma\right ) \ .
}
$$

\end